\documentclass[twocol]{ametsoc}

\journal{jcli}

\bibpunct{(}{)}{;}{a}{}{,}

\title{Faster Arctic sea ice retreat in CMIP5 than in CMIP3 due to volcanoes}

\authors{Erica Rosenblum\correspondingauthor{Scripps Institution of Oceanography, University of California at San Diego, La Jolla, CA, USA.} and Ian Eisenman}

\affiliation{Scripps Institution of Oceanography, University of California at San Diego, La Jolla, CA, USA}

\email{ejrosenb@ucsd.edu}

\abstract{The downward trend in Arctic sea ice extent is one of the most dramatic signals of climate change during recent decades. Comprehensive climate models have struggled to reproduce this, typically simulating a slower rate of sea ice retreat than has been observed. However, this bias has been widely noted to have decreased in models participating in the most recent phase of the Coupled Model Intercomparison Project (CMIP5) compared with the previous generation of models (CMIP3). Here we examine simulations from both CMIP3 and CMIP5. 
We find that simulated historical sea ice trends are influenced by volcanic forcing, which was included in all of the CMIP5 models but in only about half of the CMIP3 models. The volcanic forcing causes temporary simulated cooling in the 1980s and 1990s, which contributes to raising the simulated 1979-2013 global-mean surface temperature trends to values substantially larger than observed. We show that this warming bias is accompanied by an enhanced rate of Arctic sea ice retreat and hence a simulated sea ice trend that is closer to the observed value, which is consistent with previous findings of an approximately linear relationship between sea ice extent and global-mean surface temperature.
We find that both generations of climate models simulate Arctic sea ice that is substantially less sensitive to global warming than has been observed. The results imply that the much of the difference in Arctic sea ice trends between CMIP3 and CMIP5 occurred due to the inclusion of volcanic forcing, rather than improved sea ice physics or model resolution.
}
\begin{document}

\maketitle

\section{Introduction}

Modeling groups from around the world have contributed state-of-the-art climate model simulation results to the Coupled Model Intercomparison Project (CMIP). These simulations have natural and anthropogenic forcing from the historical period and can be compared with observations during the instrumental record to assess how well the climate models perform. The simulations are then extended to project future climate change using several different greenhouse gas concentration trajectories. There have been several CMIP phases as comprehensive climate models have continued to be developed. The two most recent phases have been the CMIP3 \citep{Meehl2007} and CMIP5 \citep{Taylor2012} ensembles, which were used to project future climate change in the Intergovernmental Panel on Climate Change (IPCC) Fourth and Fifth Assessment Reports (AR4 and AR5), respectively. 

The historical simulations have shown substantial bias in reproducing Arctic sea ice changes during the satellite record, with the models typically simulating a slower rate of sea ice retreat than has been observed \citep{Stroeve2007, Stroeve2012, Winton2011, Kay2011, Swart2015}. 
However, CMIP5 models tend to simulate faster 
sea ice trends that are more consistent with observations than CMIP3 \citep{Stroeve2012}, as illustrated in Figure 1b,c.
This has been a widely discussed feature of CMIP5, and it was highlighted in the Executive Summary of the IPCC AR5 chapter on the evaluation of climate models \citep{Flato2013}. However, the cause of this apparent improvement has remained unresolved.

Here we focus on the influence of historical volcanic forcing, which was included in all of the CMIP5 models but only about half of the CMIP3 models. Volcanic eruptions perturb the climate by injecting gases into the stratosphere that produce short-lived sulfate aerosols which reflect and absorb solar radiation. This causes rapid global surface cooling that spans approximately two to three years, which is followed by a decade-long warming period in which the climate recovers \citep{Ipcc-2013:climate}. This cooling due to volcanic eruptions has been found to be overestimated in climate models compared with observations, which can cause biases in simulated decade-scale trends \citep{Schmidt2014,Santer2014}. 
The CMIP3 models that include volcanic forcing tend to simulate global-mean surface air temperature changes that are fairly similar to the CMIP5 models, whereas CMIP3 models without volcanic forcing simulate global-mean surface air temperature changes that differ substantially from the CMIP5 simulations \citep[e.g.,][]{Schmidt2014, Knutson2013, Marotzke2015}.

Arctic sea ice extent has been found to be approximately linearly related to global-mean surface 
temperature in many of the CMIP3 and CMIP5 models \citep{Gregory2002,Winton2011,Mahlstein2012,stroeve2015}, including over periods as brief as 1979-2013 \citep{Rosenblum2016}. This implies that volcano-related biases in simulated global warming during recent decades should be associated with biases in sea ice retreat. 
Consistent with this, a number of studies have shown that volcanic forcing in climate model simulations can influence Arctic sea ice for a decade or more \citep{Stenchikov2009,Zhong2011,Zanchettin2012,Zanchettin2013,Segschneider2013,Zanchettin2014}.
Taken together,
 the results of these previous studies raise the possibility that the inclusion of volcanic forcing in all of the CMIP5 models compared to only some of the CMIP3 models could have caused a systematic change in the distribution of simulated sea ice trends during the historical period. 

We examine this effect by analyzing simulations of 1979-2013 in 118 ensemble members from 40 CMIP5 models, as well as 38 ensemble members from 19 CMIP3 models, and comparing them with observations (see details in Appendix A).
We use processed CMIP5 output from a previous study \citep{Rosenblum2016}, where we addressed whether simulated natural variability was sufficient to explain the biases in the CMIP5 ensemble-mean Arctic and Antarctic sea ice trends compared with observations.

\section{Results}
The distributions of September Arctic 
sea ice trends during 1979-2013 in the CMIP3 and CMIP5 simulations are plotted in Figure \ref{wedge}c,f. 
CMIP5 models tend to simulate a faster September Arctic sea ice retreat, which has a reduced bias compared with observations, as has been reported previously \citep{Stroeve2012,Flato2013}.
The annual-mean sea ice trend behaves similarly (Figure \ref{wedge}b,e), with the ensemble mean falling closer to the observations in CMIP5 than in CMIP3. 

It is noteworthy, however, that this decrease in bias 
in the simulated Arctic sea ice trend coincides with an increase in bias in the simulated annual-mean global-mean surface temperature trend compared to the observations during the same time period (Figure \ref{wedge}a,d). 
Although both generations of models tend to simulate too much warming, the observed global temperature trend during 1979-2013 
falls less than one standard deviation below the mean in the CMIP3 distribution, whereas the CMIP5 distribution has a larger bias (error bars in Figure \ref{wedge}d).

This warming bias is partially related to both generations of models having a tendency to simulate too much global warming during the past 10-20 years, which has been attributed to a number of factors including internal variability \citep{Ipcc-2013:climate, kosaka2013, fyfe2013}.
Additionally, the temperature trend during 1979-2013 is expected to be influenced by the eruptions of El Chich\'{o}n in 
1982 
and Pinatubo in 1991. These events cause surface cooling in the 1980s and 1990s that has been found to be overestimated in climate models \citep{Schmidt2014, Knutson2013, Marotzke2015}. Figure \ref{timeSeries}a illustrates that the large negative temperature anomalies caused by these volcanoes lead to a larger overall warming trend during this period. 
This suggests that one reason the CMIP5 ensemble-mean global warming trend during 1979-2013 is farther from the observations than in CMIP3 (Figure \ref{wedge}a,d) is because volcanic forcing is included in all of the CMIP5 models compared to about half of the CMIP3 models (Figure \ref{timeSeries}a and Table 1). 

As expected from the approximately linear relationship between sea ice cover and annual-mean global-mean surface air temperature in many CMIP3 and CMIP5 models \citep{Gregory2002,Winton2011,Mahlstein2012,stroeve2015,Rosenblum2016}, we find that the large negative temperature anomalies that are caused by simulated volcanic forcing are associated with concurrent positive sea ice cover anomalies (Figure \ref{timeSeries}). Similarly, we find that the decade-long warming periods following each eruption typically correspond with a drop in sea ice cover (Figure \ref{timeSeries}). Because these eruptions occur towards the beginning of the 1979-2013 period, they contribute to a larger overall rate of sea ice retreat (dashed-lines in Figure \ref{timeSeries}b,c). Therefore the bias in the models toward too much 1979-2013 global warming, which is elevated by volcanic forcing, appears to be associated with the larger simulated sea ice trends.

Consistent with this, we find that the major eruptions before 1979 in the CMIP3 and CMIP5 historical forcing scenarios are typically followed by a brief increase in the September and annual-mean Arctic sea ice (Figure S1).
Overall, this simulated historical Arctic sea ice response to volcanic eruptions is in agreement with previous modeling studies \citep{Segschneider2013,Zanchettin2012,Stenchikov2009,Zanchettin2014}. 

By comparing CMIP5 models with the subset of CMIP3 models that include volcanic forcing, we find that both ensembles predict a more similar distribution of both global warming trends and sea ice trends (Figure \ref{hist}a-c) than when all CMIP3 models are included (Figure \ref{wedge}d-f). Indeed, this difference in simulated volcanic forcing is typically accounted for in studies that compare simulated global warming between CMIP3 and CMIP5 \citep[e.g.,][]{Knutson2013, Watanabe2013, Knutti2012}.

Note that the influence of volcanoes is statistically significant in the CMIP3 results (red and yellow error bars in Figures \ref{hist}a-c): using the Student t-test, we can reject the null hypothesis that the two sets of CMIP3 models simulate temperature trends that are drawn from distributions with the same mean at above the 99.9\% confidence level, and the same applies to the annual and September sea ice trends. 
It should be noted by caveat that this assessment relies on the relatively small ensemble of CMIP3 models that included volcanic forcing.

\section{Discussion}

Here we examine the results presented above in the context of sea ice sensitivity to global warming \citep{Winton2011}, drawing on methods developed in a previous study \citep{Rosenblum2016}.

\subsection{Do volcanoes influence sea ice sensitivity?}

The results above suggest that volcanic forcing artificially improved simulated sea ice trends by raising the level of global warming to values larger than observed. A simple interpretation of this is that the sea ice responds to the inclusion of volcanoes just as it does if the level of global warming increases due to other factors such as greenhouse gases or internal variability. 
Here we assess this possibility by investigating whether the inclusion of volcanic forcing affects the sensitivity
of simulated sea ice cover to the level of global warming, or whether this sensitivity remains constant. 

As in \citet{Rosenblum2016}, we consider the possibility that the relationship between global warming trends and sea ice trends remains approximately constant during all 35-year periods between 1900 and 2100 (which would exactly hold if this relationship were perfectly linear). We construct two distributions of 35-year sea ice trends and associated global-mean surface air temperature trends from models that include volcanic forcing: (1) using years 1979-2013, and (2) using all available overlapping 35-year periods during 1900-2100 that are not within 10 years of a major volcanic event (i.e., Santa Maria in 1902, Agung in 1963, El Chich\'{o}n in 1982, and Pinatubo in 1991). The first distribution is meant to characterize the distribution of sea ice trends that occur under global warming including the effects of volcanic forcing, while the second characterizes the distribution of sea ice trends that occur in the same models in the absence of volcanic forcing.

In Figure \ref{gazillion}, the Arctic sea ice trend is plotted versus the annual-mean global-mean surface temperature trend, with each point representing a 35-year period in a simulation and colors representing the two distributions. By comparing the two distributions in each panel, we find that the influence of volcanic forcing has no visibly discernible impact on the sensitivity of the Arctic sea ice extent to the level of global warming. That is, for a given value on the horizontal axis in each panel of Figure \ref{gazillion}, the blue points tend to be scattered around similar vertical locations as the red points, indicating that 35-year periods that undergo similar levels of global warming to those simulated for 1979-2013 typically have similar sea ice trends, even without volcanic eruptions. This implies that the influence of volcanic forcing on simulated sea ice trends can be approximately accounted for by considering only the effect on global-mean surface temperatures.

\subsection{Comparing CMIP3 and CMIP5 sea ice sensitivities}
The relationship between global-mean surface air temperature and Arctic sea ice cover implies that biases in simulated global warming trends should be associated with biases in sea ice trends \citep{Winton2011}.
Therefore, similar to \citet{Rosenblum2016}, we examine the Arctic sea ice trend in each simulation versus the global-mean surface temperature trend.
We find that both CMIP3 models and CMIP5 models that simulate larger (hence more accurate) annual-mean sea ice trends also tend to simulate larger (hence less accurate) global warming trends (Figure \ref{effTrend}a and Table 1). 
While the CMIP3 models with volcanic forcing tend to fall in a different region of the scatter plot than those without volcanic forcing (consistent with Figure 3), the points all fall near the same line. We find similar results using September sea ice trends (Figure \ref{effTrend}b and Table 1), though this relationship appears noisier, perhaps due to a larger influence of internal variability.

We can approximately account for biases in the level of global warming by considering the Arctic ``effective sea ice trend" \citep{Rosenblum2016}, which is defined as the simulated sea ice trend scaled by the bias in simulated global warming during the same time period (where the latter is calculated as the ratio of observed to simulated annual-mean global-mean surface temperature trend; see Appendix A and \citet{Rosenblum2016} for details). The effective sea ice trend is closely related to the sea ice sensitivity \citep{Winton2011}. It provides a rough estimate of what the sea ice trend would be in each run if the observed level of global warming had been simulated.

By comparing the distributions of modeled effective Arctic sea ice trends to the observed trend, the results in Figure \ref{effTrend}c-d suggest that the modeled Arctic sea ice cover in both CMIP3 and CMIP5 would retreat far more gradually if the models simulated the observed level of global warming (see also Table 2, which includes both effective sea ice trends and sea ice sensitivities).
The effective sea ice trend in CMIP5 is slightly closer to the observations than in CMIP3, especially in September,
but the observed trend falls well outside both CMIP model distributions.

Note that this bias in simulated sea ice sensitivity is qualitatively consistent with \cite{stroeve2016}, although there are quantitative differences due to factors including the availability of CMIP5 results at the time of each analysis and differing methods used to estimate the ice sensitivity (see Appendix A). 
The possibility that simulated natural variability could explain this bias is examined in a companion paper \citep{Rosenblum2016}. 

\section{Additional Points} 

Although Southern Hemisphere sea ice cover and annual-mean global-mean surface air temperatures are also approximately linearly related in these climate models \citep{Rosenblum2016}, we find that the influence of volcanoes does not appear to have the same impact on the evolution Antarctic sea ice (Figure S2). This may be related to a range of factors, including that the aerosol forcing from both Pinatubo and El Chich\'{o}n is more concentrated in the Northern Hemisphere than the Southern Hemisphere in many datasets \citep{Arfeuille2014}, that much of the temperature change caused by volcanoes has been suggested to occur at depth in the southern ocean \citep{Fyfe2006}, that Antarctic sea ice has been suggested to only respond to supervolcanoes \citep{Zanchettin2014}, and that Antarctic sea ice extent is less correlated with annual-mean global-mean surface air temperature than Arctic sea ice extent \citep{Rosenblum2016}.

Previous studies have demonstrated that CMIP5 models simulate a smaller and more accurate climatological Arctic sea ice cover compared to CMIP3 \citep{Stroeve2012,Flato2013}. The possibility that this could be linked to sea ice trends has been considered previously, although no clear  relationship was found \citep{Massonnet2012}. Similarly, we find that the initial sea ice cover does not appear to be closely related to the sea ice trends (Figure S3). This is consistent with the approximately linear relationship between simulated Arctic sea ice cover and annual-mean global-mean surface temperatures \citep{Gregory2002,Winton2011,Mahlstein2012,stroeve2015,Rosenblum2016}. That is, if this were a perfectly linear relationship, a given amount of warming would result in the same amount of ice loss regardless of the initial amount of sea ice cover.
Note that although geographic muting effects due to the distribution of landmasses in the Arctic region \citep{Eisenman2010} can cause a departure from this linearity for very large ice extents \citep[Fig.\ S2 of][]{Armour2011}, the relationship has been found to be approximately linear for annual-mean and September ice extents similar to and smaller than modern observed values \citep[e.g., Fig.\ 2 of][]{Armour2011}).

The main results of this study are presented using sea ice extent (Figures 2-5). We find that analyzing observed and modeled sea ice area instead of extent leads to qualitatively similar results (Figures S4-S8).

Our estimate of the observed September sea ice sensitivity ($-5.67 \times 10^6$ km$^2$/K) is more than twice as large as the number reported previously by \citet{Mahlstein2012} ($-2.62 \times 10^6$ km$^2$/K), who used the ice sensitivity to make an observationally-based projection of how much  global warming it would take for the September Arctic sea ice area to decline from its 1980-1999 mean value to the nearly ice-free value of $1 \times 10^6$ km$^2$. The difference between our estimate and that in \citet{Mahlstein2012} arises due to a number of factors.
We use NASA Team sea ice extent \citep{Fetterer2002} during 1979-2013. By contrast, \citet{Mahlstein2012} use the coarser resolution Hadley Centre Sea Ice and Sea Surface Temperature (HadISST) \citep{Rayner2003} dataset,  the observed ice area rather than ice extent, and  a shorter observed time period (1980-2007). Further, they calculate the ice sensitivity using an ordinary least squares regression of ice on temperature (I.\ Mahlstein, personal communication May, 2016), which \citet{Winton2011} found to give a less accurate estimate than the method adopted here (see Appendix A).

Using CMIP3 simulations, \citet{Mahlstein2012} found that the ensemble-mean ice sensitivity during 2010-2100 was smaller than during 1980-2007 by a factor of 0.92, and hence they scaled the observed ice sensitivity by 0.92 to project the level of future global warming at which the Arctic will become nearly seasonally ice-free. 
We repeat the calculation from \citet{Mahlstein2012} using an observed ice sensitivity of $-5.67 \times 10^6$ km$^2$/K and the 1980-1999 mean September Arctic sea ice extent from the NASA Team dataset, rather than an observed sensitivity of $-2.62 \times 10^6$ km$^2$/K and the 1980-1999 mean September Arctic sea ice area from the HadISST dataset. We find that in this case the level of global warming projected to cause a nearly ice-free Arctic Ocean is approximately 1{$^\circ$}C, rather than approximately 2{$^\circ$}C as reported in \citet{Mahlstein2012}. Using NASA Team ice area rather than ice extent for the observed sensitivity and the 1980-1999 mean value yields a similar result of approximately 1{$^\circ$}C.

\section{Summary}

CMIP5 models have been found to simulate  Arctic sea ice retreat during 1979-2013 that is faster on average than in the CMIP3 models. At the same time, the CMIP5 ensemble-mean rate of global warming during 1979-2013 has been found to be larger than CMIP3. The difference in global warming has been previously attributed to historical volcanic forcing, which was included in all of the CMIP5 models but only about half of the CMIP3 models. However, the inclusion of volcanic forcing in the CMIP ensembles has not been considered, as far as the authors are aware, in previous analyses of the rate of simulated Arctic sea ice retreat. Here we show that a range of approaches 
all suggest that the change between CMIP5 and CMIP3 in the ensemble-mean 1979-2013 Arctic sea ice extent trend can also be largely attributed to the inclusion of volcanic forcing. 

Specifically, major volcanic eruptions occur during the early part of this time period, and they cause temporary cooling and ice expansion.
This exacerbates the model bias toward too much 1979-2013 global warming while reducing the model bias toward too little Arctic sea ice retreat. These results are consistent with the sea ice sensitivity not being substantially influenced by volcanic eruptions,
which would imply that the higher level of global warming caused by volcanoes should coincide with more sea ice retreat. 
This suggests that the reported improvement in simulated sea ice trends was largely an artifact of comparing simulations that had volcanic forcing with simulations that did not.
\acknowledgments
Without implying their endorsement, we are grateful to Sarah Gille, Art Miller, Paul Kushner, Neil Tandon, Fr\'{e}d\'{e}ric Lalibert\'{e}, Till Wagner, and John Fyfe for helpful comments and discussions.
The work was supported a National Science Foundation Graduate Research Fellowship and National Science Foundation grants ARC-1107795 and OCE-1357078.
Processed CMIP3 and CMIP5 data used in this study is available at http://eisenman.ucsd.edu/code.html.

\appendix[A] 
\appendixtitle{Methods}

We analyze 118 simulations of years 1979-2013 from 40 CMIP5 models \citep{Taylor2012} with historical and RCP4.5 forcing as well 38 simulations from 19 CMIP3 models \citep{Meehl2007} with historical and SRES A1B forcing. The time period we analyze is chosen based on the availability of sea ice observations at the time of analysis.
We use monthly-mean fields to compute values of global-mean surface air temperature, sea ice extent, and sea ice area. Grid cell area fields are used for models that provide them in the CMIP5 archive, and otherwise we estimate the grid cell areas based on the reported grid box vertices. For simplicity, in the distributions we treat each simulation as an ensemble member from an independent model, rather than considering which model each simulation comes from. 

CMIP3 simulations were not used in this study when either (i) temperature and sea ice data were not both available during 1979-2013 or (ii) dates reported in the file did not match the filename in the CMIP3 archive. The following CMIP3 simulations each had at least one of these issues and were excluded: all runs  of CSIRO-MK3-0; all runs of BCC-CM1; all runs of GISS-MODEL-E-H; CSIRO-MK3-5 runs 2 and 3; GISS-MODEL-E-R runs 2-9; all runs of INGIV-ECHAM; and NCAR-CCSM3.0 runs 3,4,8, and 9.
We also exclude all runs of IAP-FGOALS because the simulated sea ice extent in both hemispheres is approximately twice as large as any other CMIP3 simulation.
IPSL-CM4 reported grid cells with sea ice concentrations greater than $100\%$, which we replaced with $100\%$. Finally, note that the MRI-CGCM2-3-2a model reported having volcanic forcing in the CMIP3 documentation, but several studies found that it did not actually appear to include volcanic forcing \citep{Knutson2013,Sillmann2013}. We therefore considered this model to have not included volcanic forcing.

This study uses the processed CMIP5 values from \citet{Rosenblum2016}, where processing details are given. In the analysis of trends during years 1900-2100, we use only 80 CMIP5 simulations because 38 of the simulations do not report model output during the entirety of this longer time period.

We use observed monthly-mean sea ice extent and area from the National Snow and Ice Data Center Sea Ice Index \citep{Fetterer2002}, which uses the NASA Team algorithm. 
Missing values are filled by linearly interpolating between the same month in the previous and following years.
We use the Goddard Institute for Space Sciences Surface Temperature Analysis (GISTemp) \citep{Hansen2010} for the observed annual-mean global-mean surface temperature data.

All trends are computed using ordinary least squares regressions with time. For the sea ice sensitivity, the annual or September sea ice trend is divided by the annual global-mean surface air temperature trend. This method of estimating the sea ice sensitivity is sometimes referred to as the ``trend ratio'' \citep{Winton2011}. For the simulated effective sea ice trend, the simulated sea ice sensitivity is multiplied by the observed annual-mean global-mean surface temperature trend, as described in  \citet{Rosenblum2016}. Note that for the observations, this leads to an effective sea ice trend which is equal to the actual sea ice trend.  \citet{Winton2011} suggests that total least squares (TLS) regression between ice and temperature leads to a slightly less biased estimate of the ice sensitivity, but we find that this has a relatively small influence on the results presented here. For example, when we compute the observed Arctic sea ice sensitivity using TLS regression instead of the trend ratio, the ice sensitivity increases from $-5.67 \times 10^6$ km$^2$/K to $-5.69 \times 10^6$ km$^2$/K.

 \bibliographystyle{ametsoc2014}

\begin{thebibliography}{37}
\providecommand{\natexlab}[1]{#1}
\providecommand{\url}[1]{\texttt{#1}}
\renewcommand{\UrlFont}{\rmfamily}
\providecommand{\urlprefix}{URL }
\expandafter\ifx\csname urlstyle\endcsname\relax
  \providecommand{\doi}[1]{doi:\discretionary{}{}{}#1}\else
  \providecommand{\doi}{doi:\discretionary{}{}{}\begingroup
  \urlstyle{rm}\Url}\fi
\providecommand{\eprint}[2][]{\url{#2}}

\bibitem[{Arfeuille et~al.(2014)Arfeuille, Weisenstein, Mack, Rozanov, Peter,,
  and Br{\"{o}}nnimann}]{Arfeuille2014}
Arfeuille, F., D.~Weisenstein, H.~Mack, E.~Rozanov, T.~Peter, and
  S.~Br{\"{o}}nnimann, 2014: {Volcanic forcing for climate modeling: A new
  microphysics-based data set covering years 1600-present}. \textit{Climate of
  the Past}, \textbf{10~(1)}, 359--375, \doi{10.5194/cp-10-359-2014}.

\bibitem[{Armour et~al.(2011)Armour, Eisenman, Blanchard-Wrigglesworth,
  McCusker,, and Bitz}]{Armour2011}
Armour, K.~C., I.~Eisenman, E.~Blanchard-Wrigglesworth, K.~E. McCusker, and
  C.~M. Bitz, 2011: {The reversibility of sea ice loss in a state-of-the-art
  climate model}. \textit{Geophysical Research Letters}, \textbf{38~(16)},
  \doi{10.1029/2011GL048739}.

\bibitem[{Eisenman(2010)}]{Eisenman2010}
Eisenman, I., 2010: {Geographic muting of changes in the Arctic sea ice cover}.
  \textit{Geophysical Research Letters}, \textbf{37~(16)}, n/a--n/a,
  \doi{10.1029/2010GL043741},
  \urlprefix\url{http://doi.wiley.com/10.1029/2010GL043741}.

\bibitem[{Fetterer et~al.(2002)Fetterer, Knowles, Meier,, and
  Savoie}]{Fetterer2002}
Fetterer, F., K.~Knowles, W.~Meier, and M.~Savoie, 2002: {Sea ice index}.
  \textit{Natl. Snow and Ice Data Cent., Boulder, Colo.}, (Updated 2012.)
  http://nsidc.org/data/g02135.html.

\bibitem[{Flato et~al.(2013)}]{Flato2013}
Flato, G., and Coauthors, 2013: {Evaluation of Climate Models}. \textit{In
  Climate Change 2013: The Physical Science Basis. Contribution of Working
  Group I to the Fifth Assessment Report of the Intergovernmental Panel on
  Climate Change}, 741--866, \doi{10.1017/CBO9781107415324.020}.

\bibitem[{Fyfe(2006)}]{Fyfe2006}
Fyfe, J.~C., 2006: {Southern Ocean warming due to human influence}.
  \textit{Geophysical Research Letters}, \textbf{33}, 1--4,
  \doi{10.1029/2006GL027247}.

\bibitem[{Fyfe et~al.(2013)Fyfe, Gillett,, and Zwiers}]{fyfe2013}
Fyfe, J.~C., N.~P. Gillett, and F.~W. Zwiers, 2013: {Overestimated global
  warming over the past 20 years}. \textit{Nature Climate Change},
  \textbf{3~(9)}, 767--769, \doi{10.1038/nclimate1972}.

\bibitem[{Gregory et~al.(2002)Gregory, Stott, Cresswell, Rayner, Gordon,, and
  Sexton}]{Gregory2002}
Gregory, J.~M., P.~A. Stott, D.~J. Cresswell, N.~A. Rayner, C.~Gordon, and
  D.~M.~H. Sexton, 2002: {Recent and future changes in Arctic sea ice simulated
  by the HadCM3 AOGCM}. \textit{Geophysical Research Letters},
  \textbf{29~(24)}, 28--1--28--4, \doi{10.1029/2001GL014575}.

\bibitem[{Hansen et~al.(2010)Hansen, Ruedy, Sato,, and Lo}]{Hansen2010}
Hansen, J., R.~Ruedy, M.~Sato, and K.~Lo, 2010: {Global surface temperature
  change}. \textit{Reviews of Geophysics}, \textbf{48~(4)}, RG4004,
  \doi{10.1029/2010RG000345}.

\bibitem[{IPCC(2013)}]{Ipcc-2013:climate}
IPCC, 2013: \textit{{Climate Change 2013: The Physical Science Basis.
  Contribution of Working Group I to the Fifth Assessment Report of the
  Intergovernmental Panel on Climate Change}}, T.~Stocker, D.~Qin, G.-K.
  Plattner, M.~Tignor, S.~Allen, J.~Boschung, A.~Nauels, Y.~Xia, V.~Bex, and
  P.~Midgley, Eds., Cambridge University Press, Cambridge, UK, 1535.

\bibitem[{Kay et~al.(2011)Kay, Holland,, and Jahn}]{Kay2011}
Kay, J.~E., M.~M. Holland, and A.~Jahn, 2011: {Inter-annual to multi-decadal
  Arctic sea ice extent trends in a warming world}. \textit{Geophysical
  Research Letters}, \textbf{38~(15)}, \doi{10.1029/2011GL048008}.

\bibitem[{Knutson et~al.(2013)Knutson, Zeng,, and Wittenberg}]{Knutson2013}
Knutson, T.~R., F.~Zeng, and A.~T. Wittenberg, 2013: {Multimodel assessment of
  regional surface temperature trends: CMIP3 and CMIP5 twentieth-century
  simulations}. \textit{Journal of Climate}, \textbf{26~(22)}, 8709--8743,
  \doi{10.1175/JCLI-D-12-00567.1}.

\bibitem[{Knutti and Sedl{\'{a}}{\v{c}}ek(2012)Knutti, and
  Sedl{\'{a}}{\v{c}}ek}]{Knutti2012}
Knutti, R., and J.~Sedl{\'{a}}{\v{c}}ek, 2012: {Robustness and uncertainties in
  the new CMIP5 climate model projections}. \textit{Nature Climate Change},
  \textbf{3~(4)}, 369--373, \doi{10.1038/nclimate1716}.

\bibitem[{Kosaka and Xie(2013)Kosaka, and Xie}]{kosaka2013}
Kosaka, Y., and S.-P. Xie, 2013: {Recent global-warming hiatus tied to
  equatorial Pacific surface cooling}. \textit{Nature}, \textbf{501~(7467)},
  403--407, \doi{10.1038/nature12534}.

\bibitem[{Mahlstein and Knutti(2012)Mahlstein, and Knutti}]{Mahlstein2012}
Mahlstein, I., and R.~Knutti, 2012: {September Arctic sea ice predicted to
  disappear near 2°C global warming above present}. \textit{Journal of
  Geophysical Research}, \textbf{117~(D6)}, D06\,104,
  \doi{10.1029/2011JD016709}.

\bibitem[{Marotzke and Forster(2015)Marotzke, and Forster}]{Marotzke2015}
Marotzke, J., and P.~M. Forster, 2015: {Forcing, feedback and internal
  variability in global temperature trends}. \textit{Nature},
  \textbf{517~(7536)}, 565--570, \doi{10.1038/nature14117}.

\bibitem[{Massonnet et~al.(2012)Massonnet, Fichefet, Goosse, Bitz,
  Philippon-Berthier, Holland,, and Barriat}]{Massonnet2012}
Massonnet, F., T.~Fichefet, H.~Goosse, C.~M. Bitz, G.~Philippon-Berthier, M.~M.
  Holland, and P.-Y. Barriat, 2012: {Constraining projections of summer Arctic
  sea ice}. \textit{The Cryosphere}, \textbf{6~(6)}, 1383--1394,
  \doi{10.5194/tc-6-1383-2012}.

\bibitem[{Meehl et~al.(2007)Meehl, Covey, Delworth, Latif, McAvaney, Mitchell,
  Stouffer,, and Taylor}]{Meehl2007}
Meehl, G.~A., C.~Covey, T.~Delworth, M.~Latif, B.~McAvaney, J.~F.~B. Mitchell,
  R.~J. Stouffer, and K.~E. Taylor, 2007: {The WCRP CMIP3 multimodel dataset: A
  new era in climatic change research}. \textit{Bulletin of the American
  Meteorological Society}, \textbf{88~(9)}, 1383--1394,
  \doi{10.1175/BAMS-88-9-1383}.

\bibitem[{Rayner et~al.(2003)Rayner, Parker, Horton, Folland, Alexander,
  Rowell, Kent,, and Kaplan}]{Rayner2003}
Rayner, N.~A., D.~Parker, E.~Horton, C.~Folland, L.~Alexander, D.~Rowell,
  E.~Kent, and A.~Kaplan, 2003: {Global analyses of sea surface temperature,
  sea ice, and night marine air temperature since the late nineteenth century}.
  \textit{Journal of Geophysical Research}, \textbf{108~(D14)}, 4407,
  \doi{10.1029/2002JD002670}.

\bibitem[{Rosenblum and Eisenman(2016)Rosenblum, and Eisenman}]{Rosenblum2016}
Rosenblum, E., and I.~Eisenman, 2016: Can natural variability explain the
  difference between observed and modeled sea ice trends? Submitted,
  http://eisenman.ucsd.edu/papers/Rosenblum--Eisenman--2016a.pdf.

\bibitem[{Santer et~al.(2014)}]{Santer2014}
Santer, B.~D., and Coauthors, 2014: {Volcanic contribution to decadal changes
  in tropospheric temperature}. \textit{Nature Geoscience}, \textbf{7~(3)},
  185--189, \doi{10.1038/ngeo2098}.

\bibitem[{Schmidt et~al.(2014)Schmidt, Shindell,, and Tsigaridis}]{Schmidt2014}
Schmidt, G.~A., D.~T. Shindell, and K.~Tsigaridis, 2014: {Reconciling warming
  trends}. \textit{Nature Geoscience}, \textbf{7~(3)}, 158--160,
  \doi{10.1038/ngeo2105}.

\bibitem[{Segschneider et~al.(2013)}]{Segschneider2013}
Segschneider, J., and Coauthors, 2013: {Impact of an extremely large magnitude
  volcanic eruption on the global climate and carbon cycle estimated from
  ensemble Earth System Model simulations}. \textit{Biogeosciences},
  \textbf{10}, 669--687, \doi{10.5194/bg-10-669-2013}.

\bibitem[{Sillmann et~al.(2013)Sillmann, Kharin, Zhang, Zwiers,, and
  Bronaugh}]{Sillmann2013}
Sillmann, J., V.~V. Kharin, X.~Zhang, F.~W. Zwiers, and D.~Bronaugh, 2013:
  {Climate extremes indices in the CMIP5 multimodel ensemble: Part 1. Model
  evaluation in the present climate}. \textit{Journal of Geophysical Research:
  Atmospheres}, \textbf{118~(4)}, 1716--1733, \doi{10.1002/jgrd.50203}.

\bibitem[{Stenchikov et~al.(2009)Stenchikov, Delworth, Ramaswamy, Stouffer,
  Wittenberg,, and Zeng}]{Stenchikov2009}
Stenchikov, G., T.~L. Delworth, V.~Ramaswamy, R.~J. Stouffer, A.~Wittenberg,
  and F.~Zeng, 2009: {Volcanic signals in oceans}. \textit{Journal of
  Geophysical Research: Atmospheres}, \textbf{114~(16)}, 1--13,
  \doi{10.1029/2008JD011673}.

\bibitem[{Stroeve et~al.(2007)Stroeve, Holland, Meier, Scambos,, and
  Serreze}]{Stroeve2007}
Stroeve, J., M.~M. Holland, W.~Meier, T.~Scambos, and M.~Serreze, 2007: {Arctic
  sea ice decline: Faster than forecast}. \textit{Geophysical Research
  Letters}, \textbf{34~(9)}, L09\,501, \doi{10.1029/2007GL029703}.

\bibitem[{Stroeve and Notz(2015)Stroeve, and Notz}]{stroeve2015}
Stroeve, J., and D.~Notz, 2015: {Insights on past and future sea-ice evolution
  from combining observations and models}. \textit{Global and Planetary
  Change}, \textbf{135}, 119--132, \doi{10.1016/j.gloplacha.2015.10.011}.

\bibitem[{Stroeve and Notz(2016)Stroeve, and Notz}]{stroeve2016}
Stroeve, J., and D.~Notz, 2016: {Corrigendum to insights on past and future
  sea-ice evolution from combining observations and models [Glob. Planet.
  Change (2015) 119-132]}. \textit{Global and Planetary Change}, \textbf{144},
  270, \doi{10.1016/j.gloplacha.2016.07.003}.

\bibitem[{Stroeve et~al.(2012)Stroeve, Kattsov, Barrett, Serreze, Pavlova,
  Holland,, and Meier}]{Stroeve2012}
Stroeve, J.~C., V.~Kattsov, A.~Barrett, M.~Serreze, T.~Pavlova, M.~Holland, and
  W.~N. Meier, 2012: {Trends in Arctic sea ice extent from CMIP5, CMIP3 and
  observations}. \textit{Geophysical Research Letters}, \textbf{39~(16)},
  L16\,502, \doi{10.1029/2012GL052676}.

\bibitem[{Swart et~al.(2015)Swart, Fyfe, Hawkins, Kay,, and Jahn}]{Swart2015}
Swart, N.~C., J.~C. Fyfe, E.~Hawkins, J.~E. Kay, and A.~Jahn, 2015: {Influence
  of internal variability on Arctic sea-ice trends}. \textit{Nature Climate
  Change}, \textbf{5~(2)}, 86--89, \doi{10.1038/nclimate2483}.

\bibitem[{Taylor et~al.(2012)Taylor, Stouffer,, and Meehl}]{Taylor2012}
Taylor, K.~E., R.~J. Stouffer, and G.~A. Meehl, 2012: {An overview of CMIP5 and
  the experiment design}. \textit{Bulletin of the American Meteorological
  Society}, \textbf{93~(4)}, 485--498, \doi{10.1175/BAMS-D-11-00094.1}.

\bibitem[{Watanabe et~al.(2013)Watanabe, Kamae, Yoshimori, Oka, Sato, Ishii,
  Mochizuki,, and Kimoto}]{Watanabe2013}
Watanabe, M., Y.~Kamae, M.~Yoshimori, A.~Oka, M.~Sato, M.~Ishii, T.~Mochizuki,
  and M.~Kimoto, 2013: {Strengthening of ocean heat uptake efficiency
  associated with the recent climate hiatus}. \textit{Geophysical Research
  Letters}, \textbf{40~(12)}, 3175--3179, \doi{10.1002/grl.50541}.

\bibitem[{Winton(2011)}]{Winton2011}
Winton, M., 2011: {Do Climate Models Underestimate the Sensitivity of Northern
  Hemisphere Sea Ice Cover?} \textit{Journal of Climate}, \textbf{24~(15)},
  3924--3934, \doi{10.1175/2011JCLI4146.1}.

\bibitem[{Zanchettin et~al.(2013)Zanchettin, Bothe, Graf, Lorenz, Luterbacher,
  Timmreck,, and Jungclaus}]{Zanchettin2013}
Zanchettin, D., O.~Bothe, H.~F. Graf, S.~J. Lorenz, J.~Luterbacher,
  C.~Timmreck, and J.~H. Jungclaus, 2013: {Background conditions influence the
  decadal climate response to strong volcanic eruptions}. \textit{Journal of
  Geophysical Research: Atmospheres}, \textbf{118~(10)}, 4090--4106,
  \doi{10.1002/jgrd.50229}.

\bibitem[{Zanchettin et~al.(2014)Zanchettin, Bothe, Timmreck, Bader, Beitsch,
  Graf, Notz,, and Jungclaus}]{Zanchettin2014}
Zanchettin, D., O.~Bothe, C.~Timmreck, J.~Bader, A.~Beitsch, H.-F. Graf,
  D.~Notz, and J.~H. Jungclaus, 2014: {Inter-hemispheric asymmetry in the
  sea-ice response to volcanic forcing simulated by MPI-ESM (COSMOS-Mill)}.
  \textit{Earth System Dynamics}, \textbf{5~(1)}, 223--242,
  \doi{10.5194/esd-5-223-2014}.

\bibitem[{Zanchettin et~al.(2012)Zanchettin, Timmreck, Graf, Rubino, Lorenz,
  Lohmann, Kr{\"{u}}ger,, and Jungclaus}]{Zanchettin2012}
Zanchettin, D., C.~Timmreck, H.-F. Graf, A.~Rubino, S.~Lorenz, K.~Lohmann,
  K.~Kr{\"{u}}ger, and J.~H. Jungclaus, 2012: {Bi-decadal variability excited
  in the coupled ocean–atmosphere system by strong tropical volcanic
  eruptions}. \textit{Climate Dynamics}, \textbf{39~(1-2)}, 419--444,
  \doi{10.1007/s00382-011-1167-1}.

\bibitem[{Zhong et~al.(2011)Zhong, Miller, Otto-Bliesner, Holland, Bailey,
  Schneider,, and Geirsdottir}]{Zhong2011}
Zhong, Y., G.~H. Miller, B.~L. Otto-Bliesner, M.~M. Holland, D.~A. Bailey,
  D.~P. Schneider, and A.~Geirsdottir, 2011: {Centennial-scale climate change
  from decadally-paced explosive volcanism: a coupled sea ice-ocean mechanism}.
  \textit{Climate Dynamics}, \textbf{37~(11-12)}, 2373--2387,
  \doi{10.1007/s00382-010-0967-z}.

\end{thebibliography}

\renewcommand{\thetable}{\arabic{table}}

\begin{table*} [t]
\centering  
\begin{tabular}{l c c c c} 
\hline\hline   
Models w/ volc & Number of Simulations  & Ann. global warming  & Ann. sea ice trend & Sept. sea ice trend \\ [0.5ex] 
\hline                  
gfdl-cm2-0 			& 1 &  0.27 & -0.72 &  	-0.59 	\\ [1ex]      
gfdl-cm2-1 			& 1 &  0.28 & -0.51 &  	-0.68		\\ [1ex]      
giss-model-e-r 		& 1 &  0.20 & -0.14 &  	-0.20	\\ [1ex]      
miroc3-2-hires 		& 1 &  0.34 & -0.54 &  	-0.73 		\\ [1ex]      
miroc3-2-medres 	& 3 &  0.20 & -0.26 & 	-0.32	\\ [1ex]      
miub-echo-g 		& 3 &  0.21 & -0.28 &  	-0.34		\\ [1ex]      
ncar-ccsm3-0 		& 5 &  0.29 & -0.45 &  	-0.60		\\ [1ex]      
ukmo-hadgem1 		& 1 &  0.25 & -0.50 & 	-0.67		\\ [1ex]      
all models w/ volc  & 16 & 0.25 & -0.39 &	-0.49	\\ [1ex]      
\hline 
\hline 
Models w/o volc & Number of Simulations  & Ann. global warming  & Ann. sea ice trend & Sept. sea ice trend \\ [0.5ex] 
\hline                  
bccr-bcm2-0 		& 1 & 0.14  & -0.27 &  -0.35    \\ [1ex]      
cccma-cgcm3-1 		& 5 & 0.24  & -0.15 &  -0.18    \\ [1ex]      
cccma-cgcm3-1-t63 	& 1 & 0.29  & -0.23 &  -0.24    \\ [1ex]      
cnrm-cm3 			& 1 & 0.17  &  0.02 &  -0.25    \\ [1ex]      
csiro-mk3-5 		& 1 & 0.21  & -0.15 &  -0.26    \\ [1ex]      
giss-aom 			& 2 & 0.14  & -0.17 &  -0.23    \\ [1ex]      
inmcm3-0 			& 1 & 0.26  & -0.40 &  -0.53    \\ [1ex]      
ipsl-cm4 			& 1 & 0.28  & -0.49 &  -0.58    \\ [1ex]      
mpi-echam5 			& 3 & 0.15  & -0.21 &  -0.22    \\ [1ex]      
mri-cgcm2-3-2a 		& 5 & 0.13  & -0.10 &  -0.11    \\ [1ex]      
ukmo-hadcm3 		& 1 & 0.16  & -0.20 &  -0.29    \\ [1ex]      
all models w/o volc & 22 & 0.18 & -0.18 &  -0.23   \\ [1ex]      
\hline 
\end{tabular}
\label{parameters} 
\caption{For each CMIP3 model, the number of runs, 1979-2013 annual-mean global-mean surface temperature trend (K/decade) averaged over the runs, and 1979-2013 annual-mean and September Arctic sea ice trends ($10^6$ km$^2$/decade) averaged over the runs.
See http://www-pcmdi.llnl.gov/ipcc/model\_documentation/ipcc\_model\_documentation.php for a list of the modeling centers associated with each CMIP3 model listed here. Note that similar information for the CMIP5 models is given in Table 1 of \citet{Rosenblum2016}.}
\end{table*}

\begin{table*} [t]
\centering  
\begin{tabular}{l c c} 
\hline\hline   
 &Ann. effective ice trend  & Sept. effective ice trend  \\ [0.5ex] 
\hline                  
CMIP3 & -0.19 (0.11) & -0.24 (0.12)      \\ [1ex]      
CMIP5 & -0.23 (0.09) & -0.36 (0.17)      \\ [1ex]      
Observations & -0.53 & -0.89    \\ [1ex]      
\hline 
\hline 
&Ann. ice sensitivity  & Sept. ice sensitivity \\ [0.5ex] 
\hline                  
CMIP3 & -1.23 (0.69) & -1.53 (0.75)      \\ [1ex]      
CMIP5 & -1.46 (0.56) & -2.29 (1.10)      \\ [1ex]      
Observations & -3.40 & -5.67     \\ [1ex]      
\hline
\end{tabular}
\label{parameters2} 
\caption{Observed as well as CMIP3 and CMIP5 ensemble-mean effective sea ice trends ($10^6$ km$^2$/decade) 
and ice sensitivity ($10^6$ km$^2$/K), as shown in Figure \ref{effTrend}. The standard deviations among the ensemble members are indicated in parentheses. Note that the ice sensitivity is equal to the effective ice trend divided by the observed temperature trend, which is 0.16 K/decade.}
\end{table*}

\renewcommand{\thefigure}{\arabic{figure}}

\begin{figure*}[t]
 \centering
 \includegraphics[width=183mm]{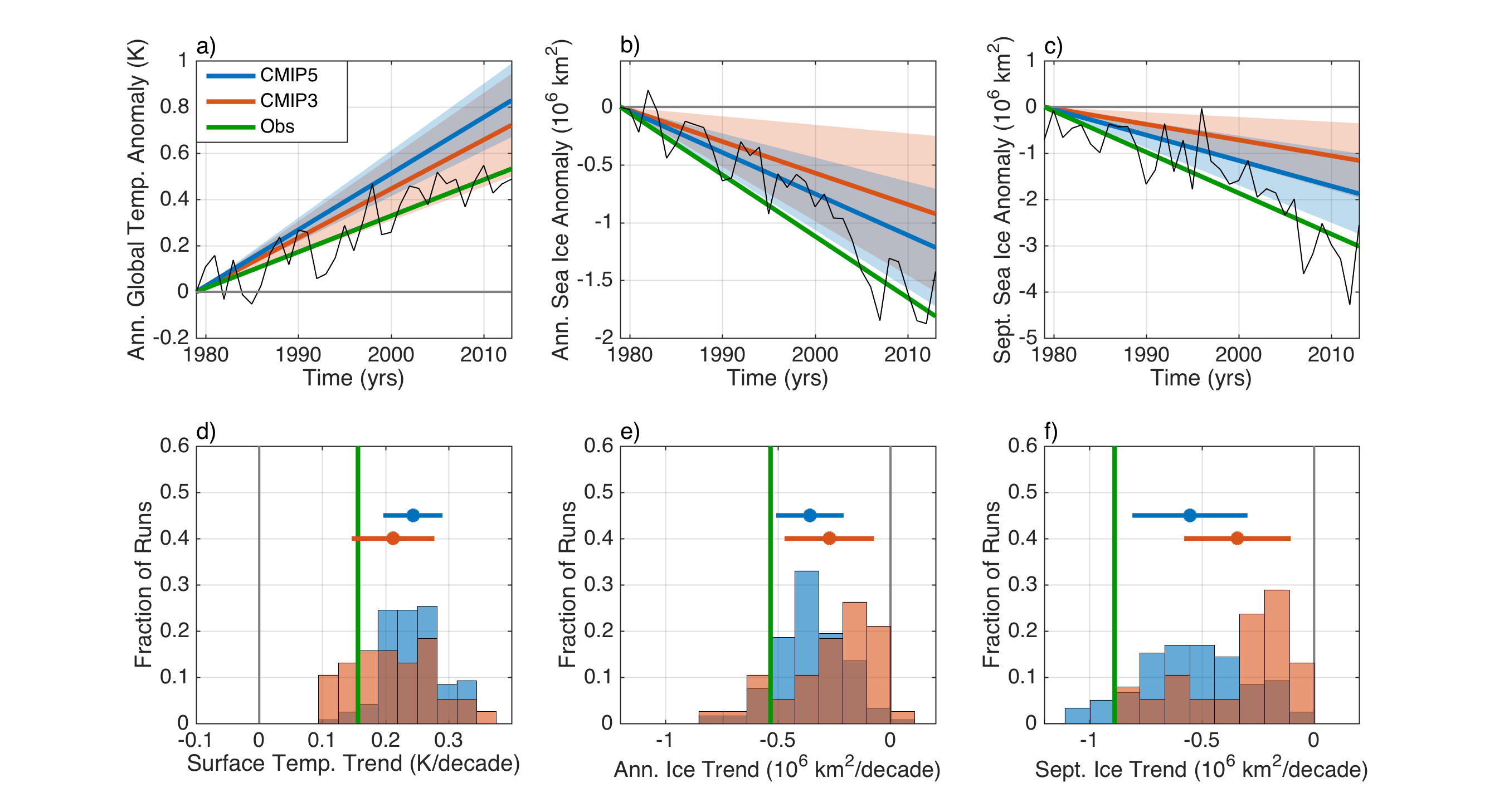}
  \caption{Observed as well as CMIP3 and CMIP5 modeled trends in (a,c) annual-mean global-mean surface temperature, (b,e) annual-mean Arctic sea ice extent, and (c,f) September Arctic sea ice extent. (a-c) Here the trends are illustrated as straight lines indicating the anomaly from 1979, and shadings indicate one standard deviation among the CMIP3 or CMIP5 trends around the ensemble means. The observed time series is also included (black, shifted vertically so linear trend goes through zero in 1979). 
(d-f) Histograms illustrating the distributions of CMIP3 and CMIP5 trends. Standard deviations among the distributions around the ensemble means are indicated by blue and red error bars above the distributions, and the observed trends are indicated by vertical green lines.}
\label{wedge}
\end{figure*}

\begin{figure*}[t]
 \centering
 \includegraphics[width=183mm]{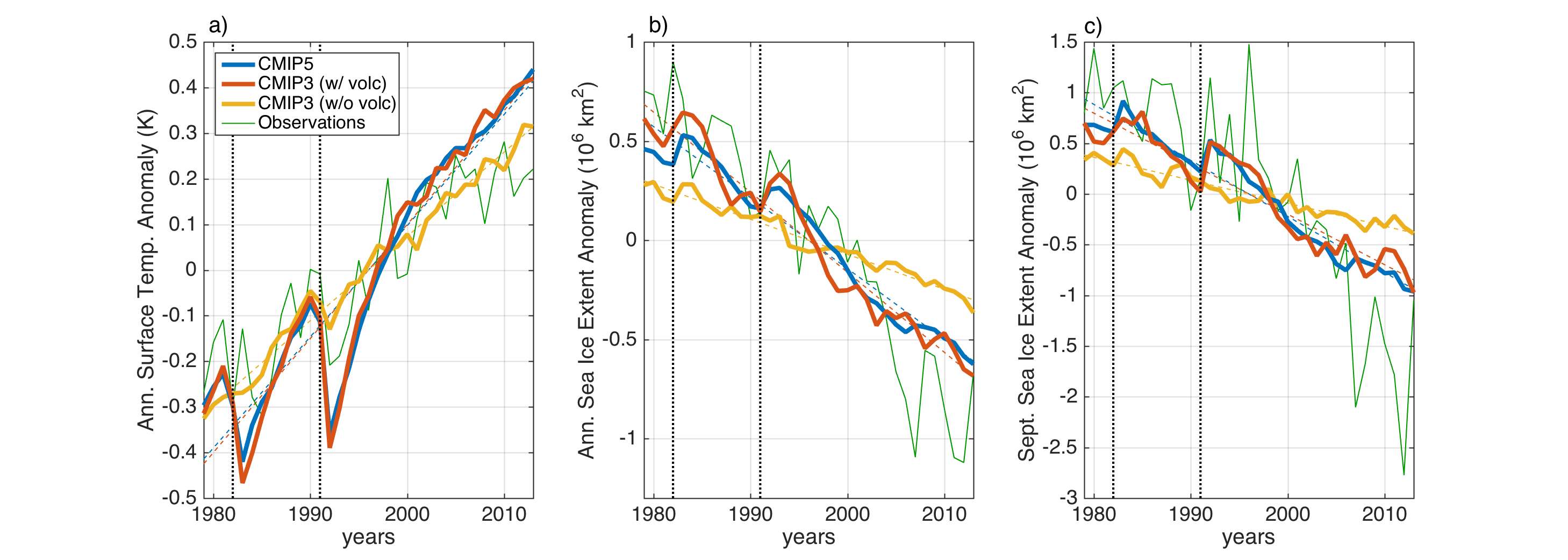}
 \caption{Observed and modeled (a) annual-mean global-mean surface temperature, (b) annual-mean Arctic sea ice area, and (c) September Arctic sea ice area. Anomalies from the average value during the plotted time period are shown for the observations (green), the CMIP5 ensemble mean (blue), and the ensemble mean of CMIP3 models with (red) and without (yellow) volcanic forcing. The linear trend associated with each time series is also indicated (dashed lines). This figure illustrates how the cooling effects associated with the eruptions of El Chich\'{o}n (1982) and Pinatubo (1991) (vertical dotted lines) result in a faster global-mean temperature trends and sea ice cover trends.}
\label{timeSeries}
\end{figure*}

\begin{figure*}[t]
 \centering
 \includegraphics[width=183mm]{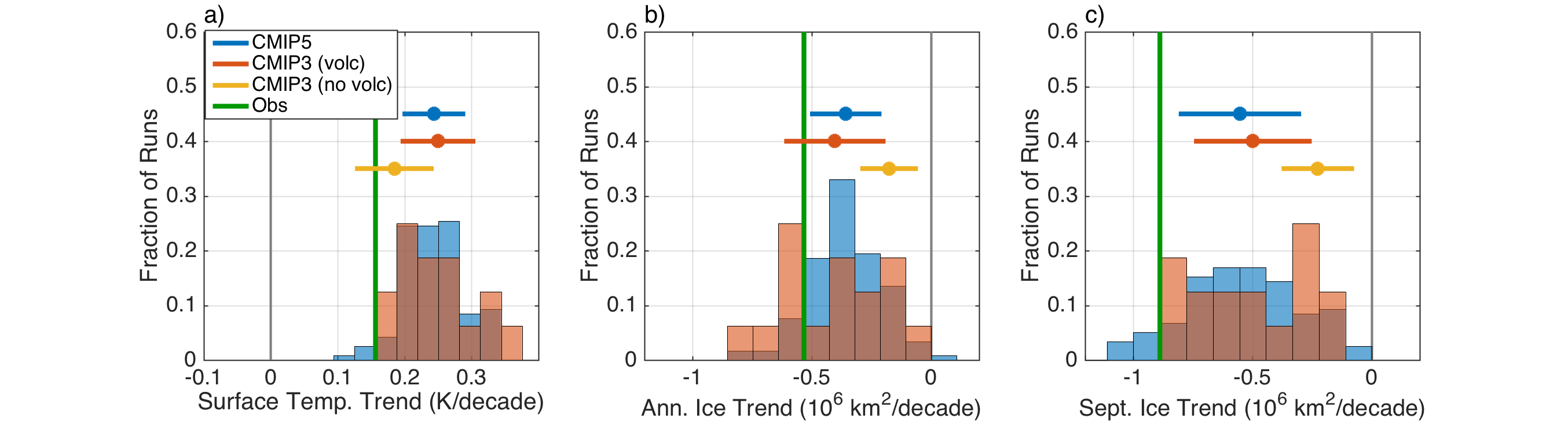}
 \caption{As in Figure \ref{wedge}d-f, but neglecting the CMIP3 simulations that do not include volcanic forcing. The ensemble mean and standard deviation of the CMIP3 simulations that do not include volcanic forcing are also indicated (yellow error bar).
}
\label{hist}
\end{figure*}

\begin{figure*}[t] 
 \centering
 \includegraphics[width=183mm]{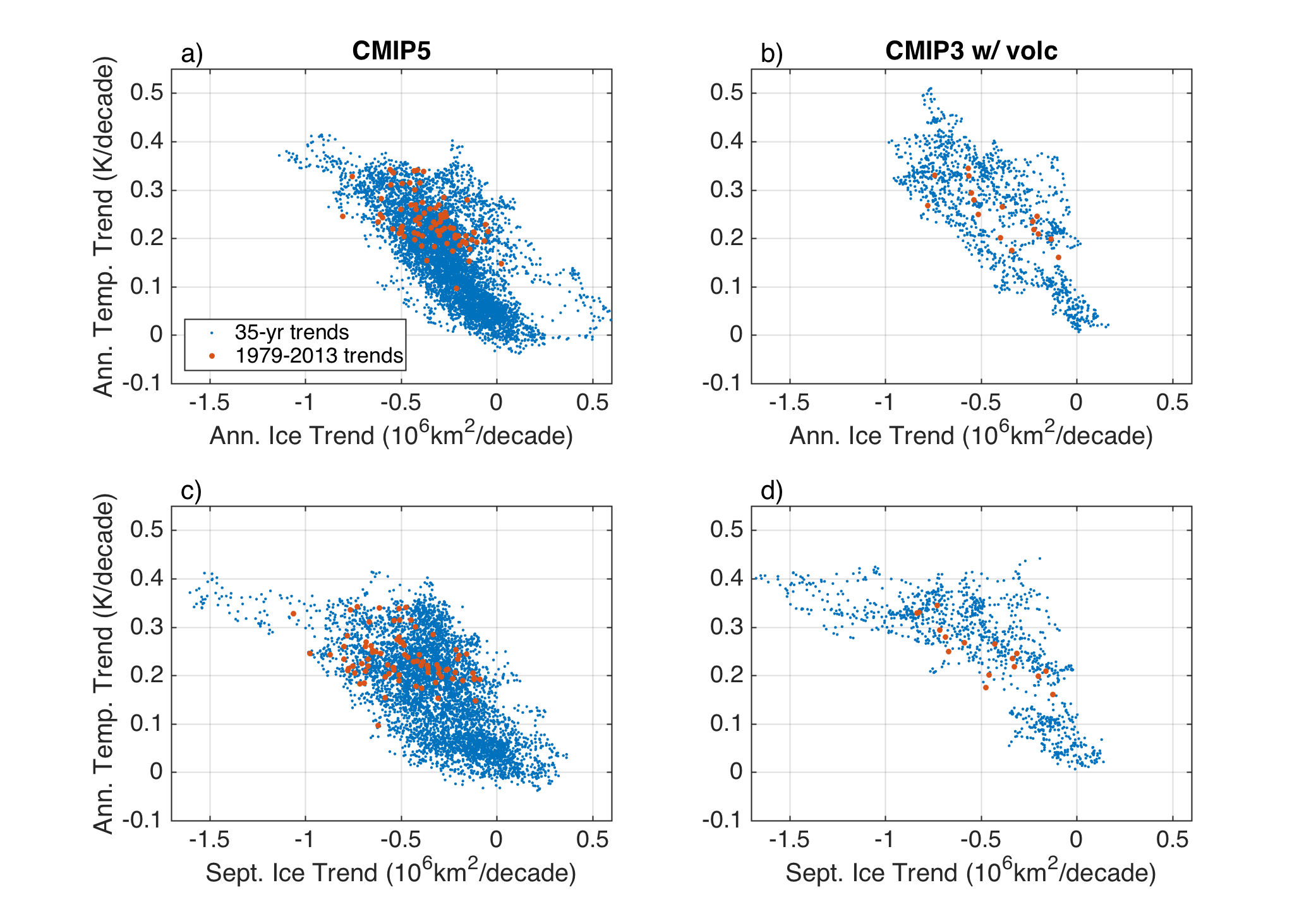}
 \caption{Simulated 35-year annual-mean global-mean temperature trends plotted versus the corresponding Arctic sea ice trends.
Annual-mean sea ice trends are plotted in the top row, and September sea ice trends are plotted in the bottom row; CMIP5 models are plotted in the left column and CMIP3 models that included volcanic forcing are plotted in the right column. Trends from 1979-2013 are indicated in red, and all available 35-year time periods between 1900-2100 that were not within 10 years of a major volcanic event are indicated in blue. The major volcanic events are Santa Maria in 1902, Agung in 1963, El Chich\'{o}n in 1982, and Pinatubo in 1991.}
\label{gazillion}
\end{figure*}

\begin{figure*}[t]
 \centering
 \includegraphics[width=183mm]{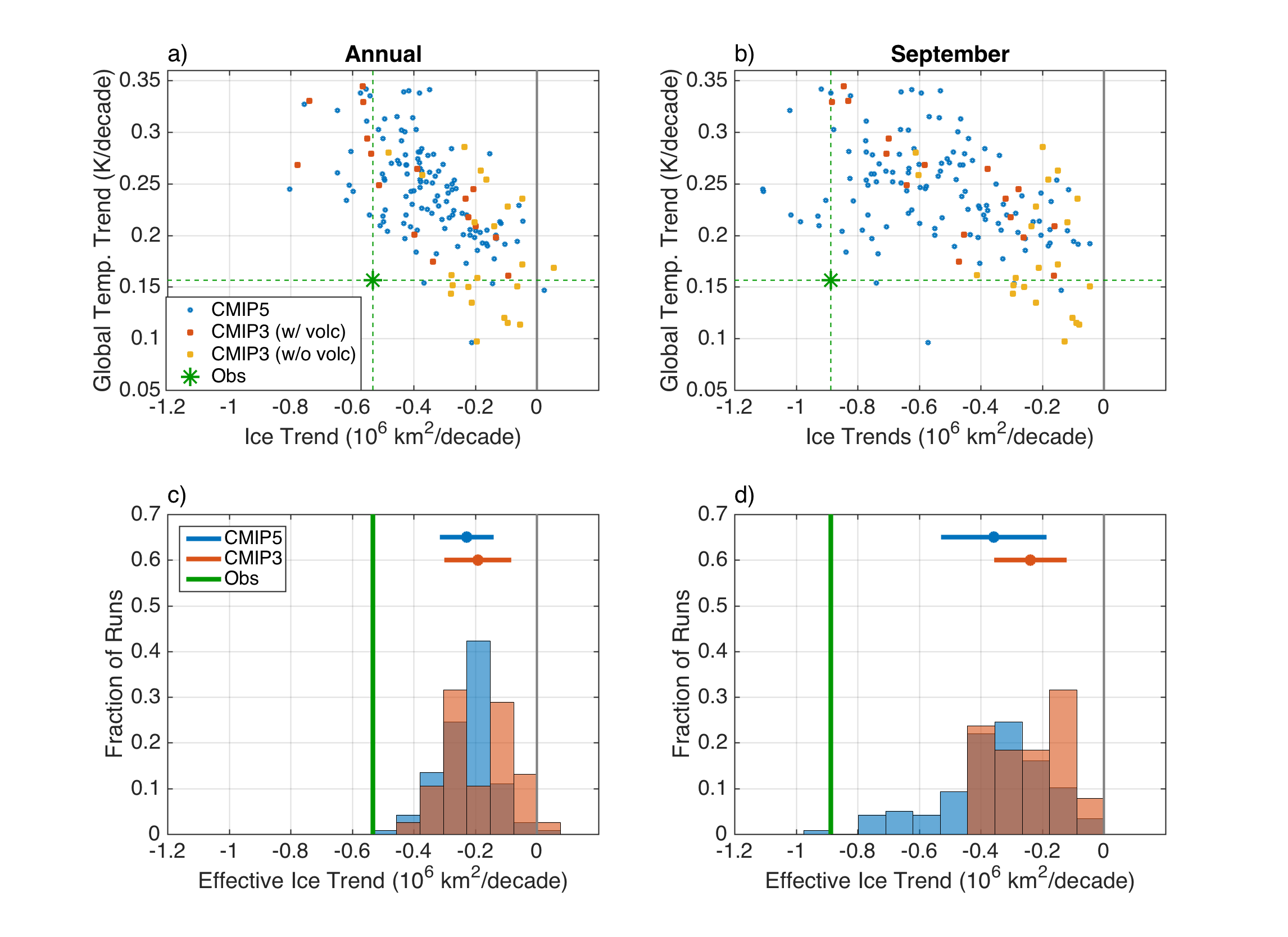}
 \caption{Observed and modeled annual-mean global-mean temperature trends plotted versus the corresponding Arctic (a) annual-mean sea ice trends and (b) September sea ice trends. CMIP5 models (blue), CMIP3 models with volcanic forcing (red), and CMIP3 models without volcanic forcing (yellow) are plotted, and dashed green lines represent the observed trend. The histograms show the Arctic (c) annual-mean effective sea ice trends and (d) September effective sea ice trends (see text for details), with the observed trend indicated by a thick green line. 
Standard deviations of the distributions around the ensemble means are also indicated. 
Note that the histograms in Fig. 1b,c describe the distributions of horizontal coordinate values in Fig. 5a,b.}
\label{effTrend}
\end{figure*}

\end{document}


\renewcommand{\thefigure}{S\arabic{figure}}

\begin{figure*}[!h]
 \centering

\bigskip

{\large \bf Supplemental Material for ``Faster Arctic sea ice retreat in CMIP5 than in CMIP3 due to volcanoes"}

\bigskip

{\sc Erica Rosenblum and Ian Eisenman}

\vspace{1in}

   \includegraphics[width=183mm]{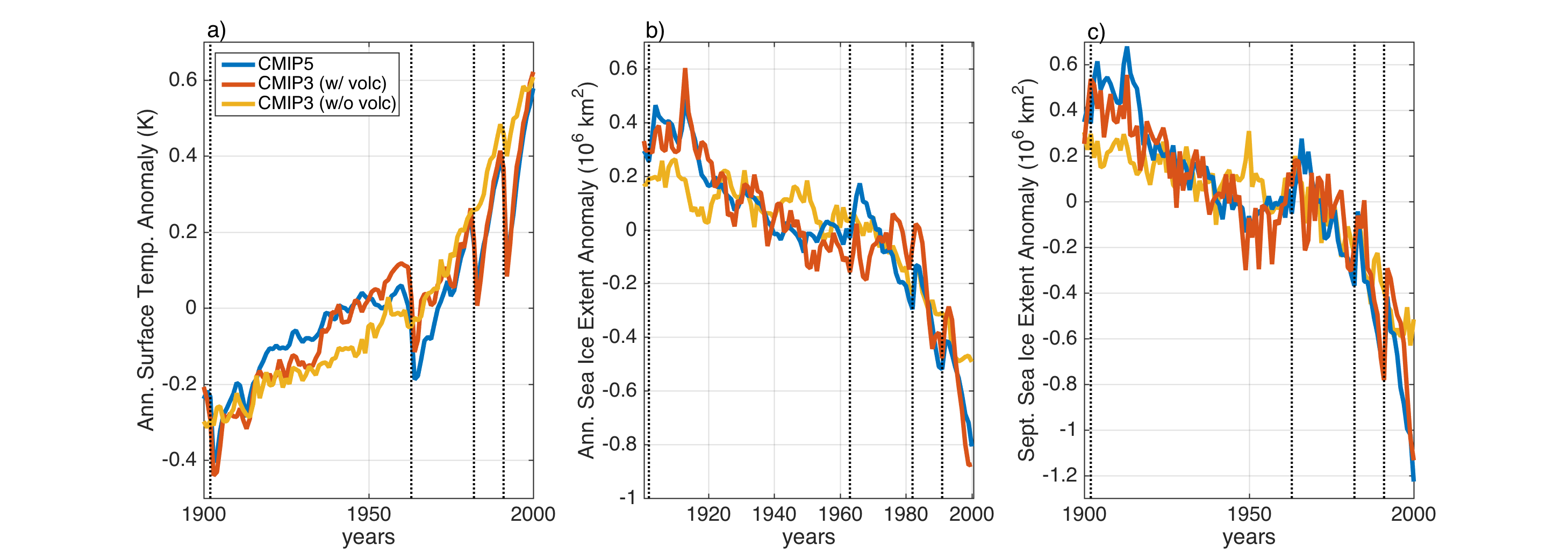}
 \caption{As in Figure 2, but showing the simulations only and looking at an extended time period that includes the additional volcanic eruptions of Santa Maria (1901) and Agung (1963).}
\label{timeSeriesLong}
\end{figure*}

\begin{figure*}
 \centering
   \includegraphics[width=183mm]{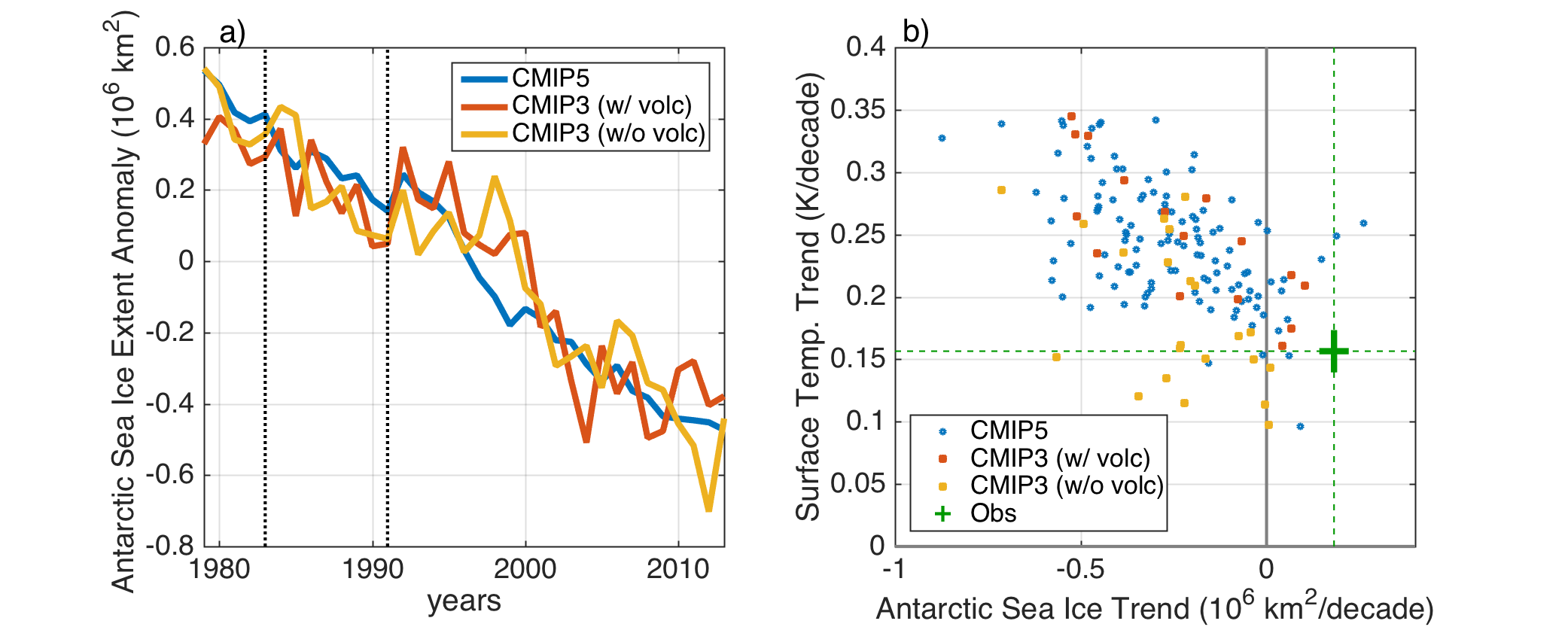}
 \caption{
 As in (a) Figure 2b and (b) Figure 3a, but for annual-mean Antarctic sea ice trends. 
 }
\label{Antarctic}
\end{figure*}

\begin{figure*}
 \centering
   \includegraphics[width=183mm]{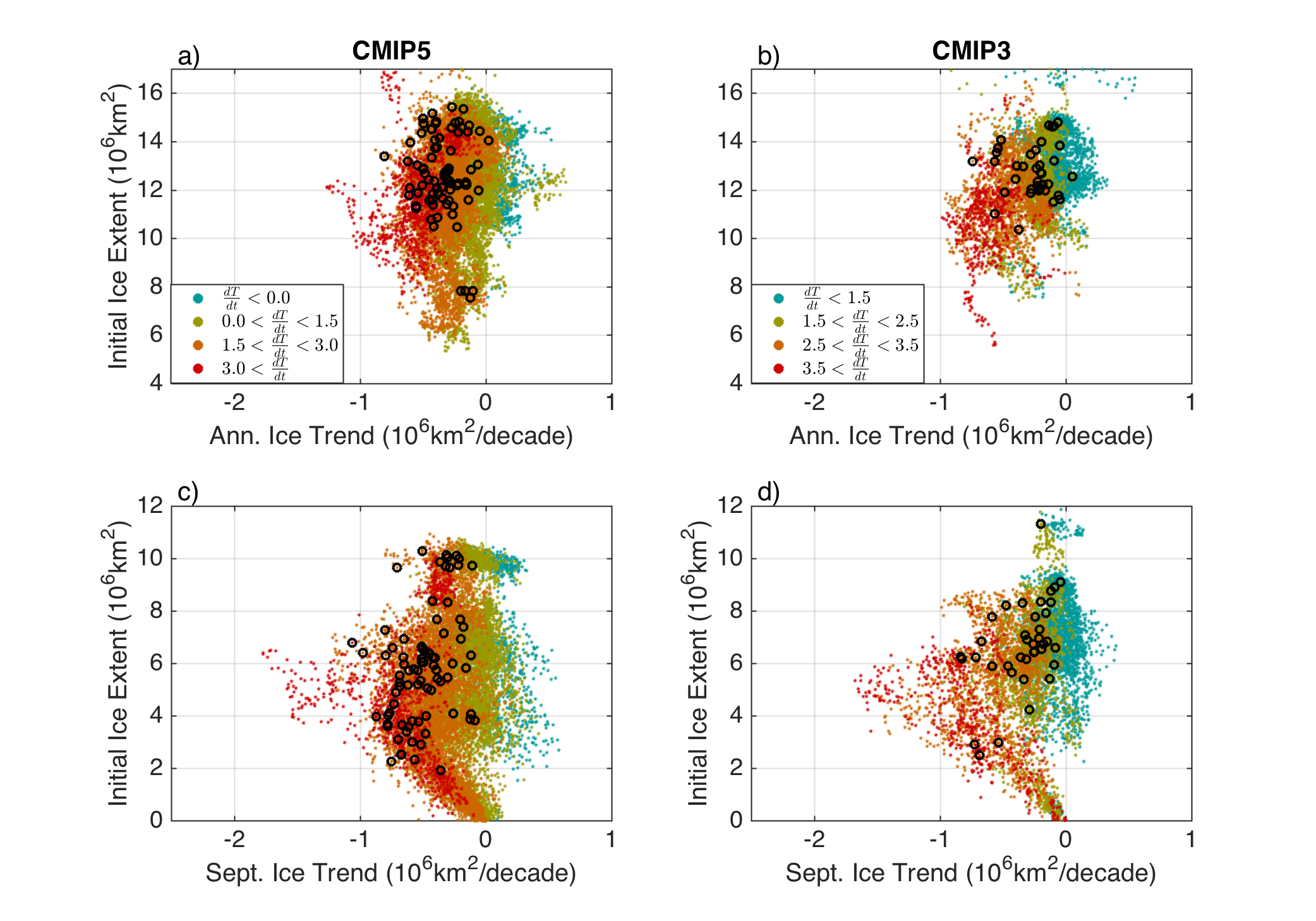}

 \caption{As in Figure 4 but using initial Arctic sea ice extent on the horizontal axis rather than global surface temperature trends. Additionally, the level of warming is indicated by the colors. 1979-2013 trends are indicated in black.}
\label{initExtent}
\end{figure*}

\begin{figure*}
 \centering
  \includegraphics[width=183mm]{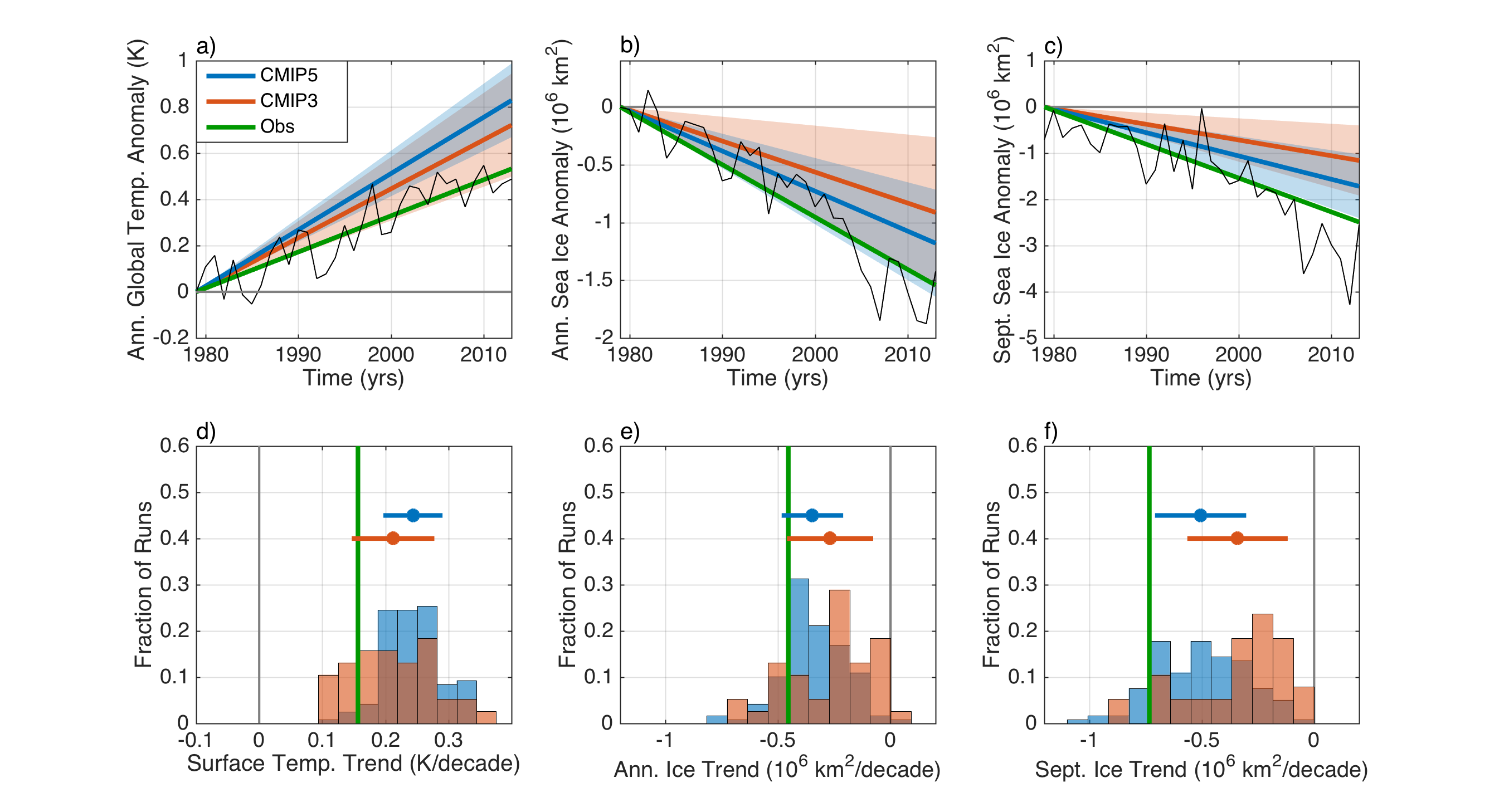}
\caption{As in Figure 1, but using sea ice area rather than extent.}
\label{histArea}
\end{figure*}

\begin{figure*}
 \centering
  \includegraphics[width=183mm]{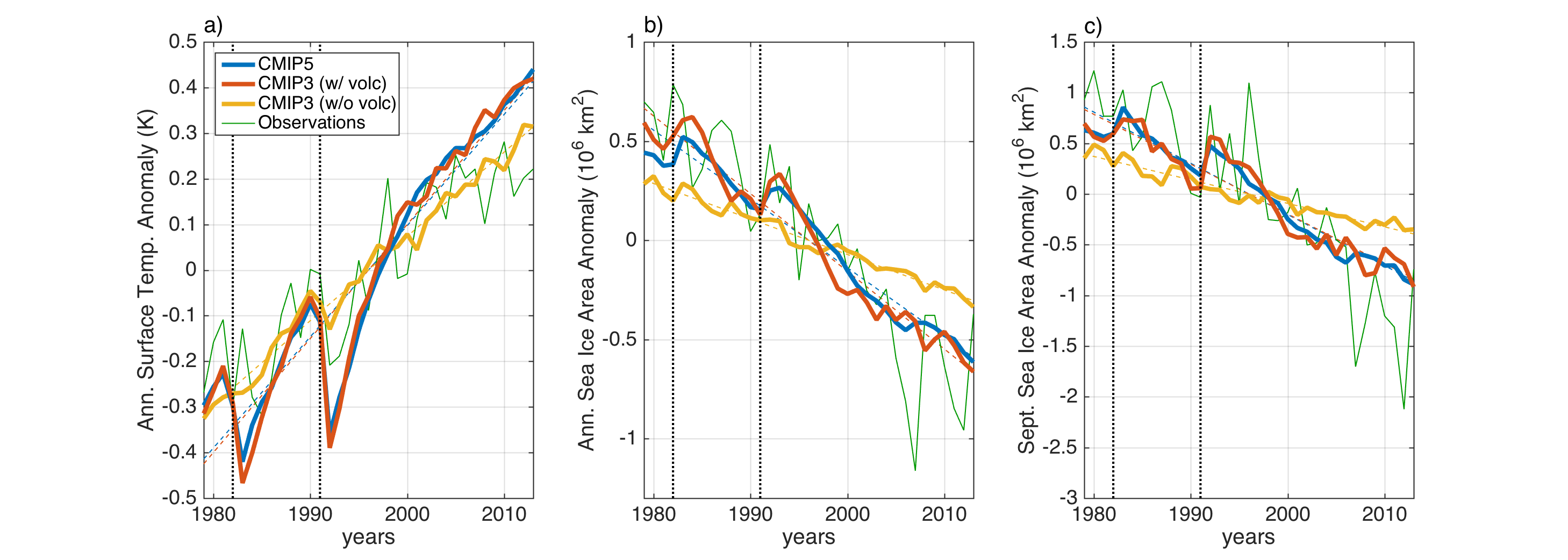}
 \caption{As in Figure 2, but using sea ice area rather than extent.}
\label{timeSeriesArea}
\end{figure*}

\begin{figure*}
 \centering
  \includegraphics[width=183mm]{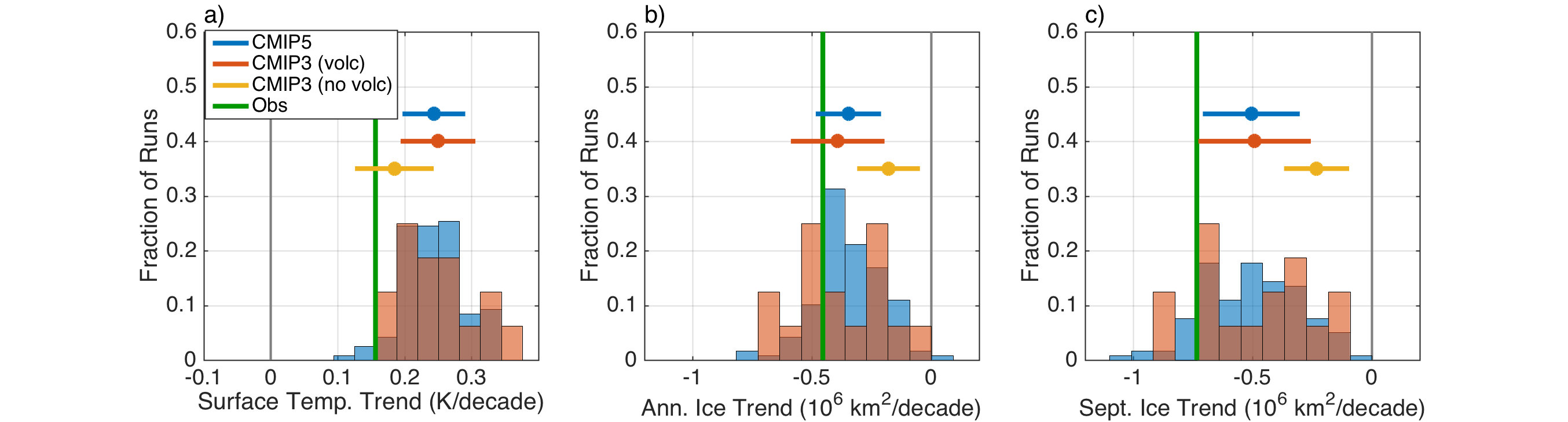}
\caption{As in Figure 3, but using sea ice area rather than extent.}
\label{histArea2}
\end{figure*}

\begin{figure*}
 \centering
  \includegraphics[width=183mm]{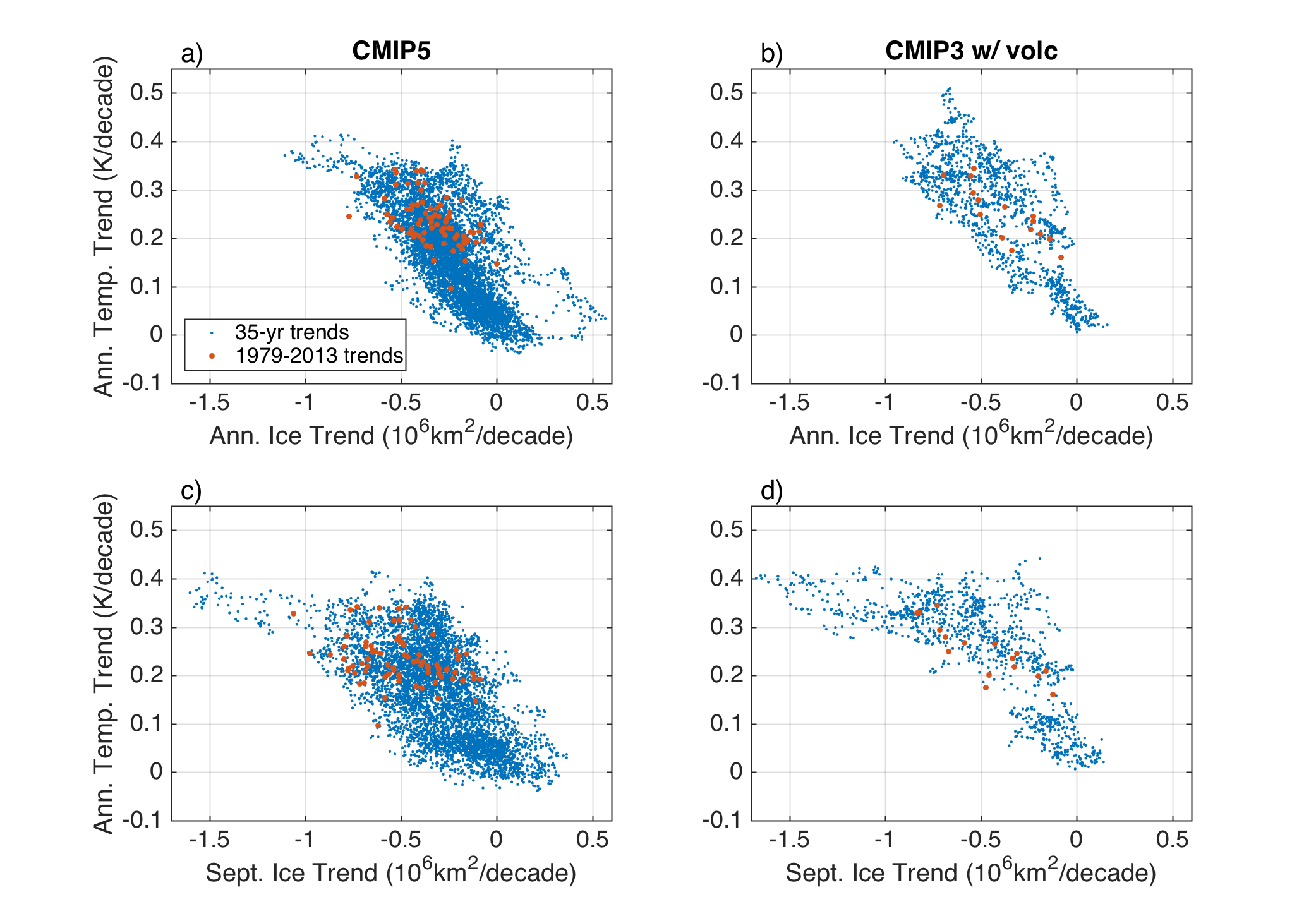}
 \caption{As in Figure 4, but using sea ice area rather than extent.}
 \label{gazillionArea}
\end{figure*}

\begin{figure*}
 \centering
    \includegraphics[width=183mm]{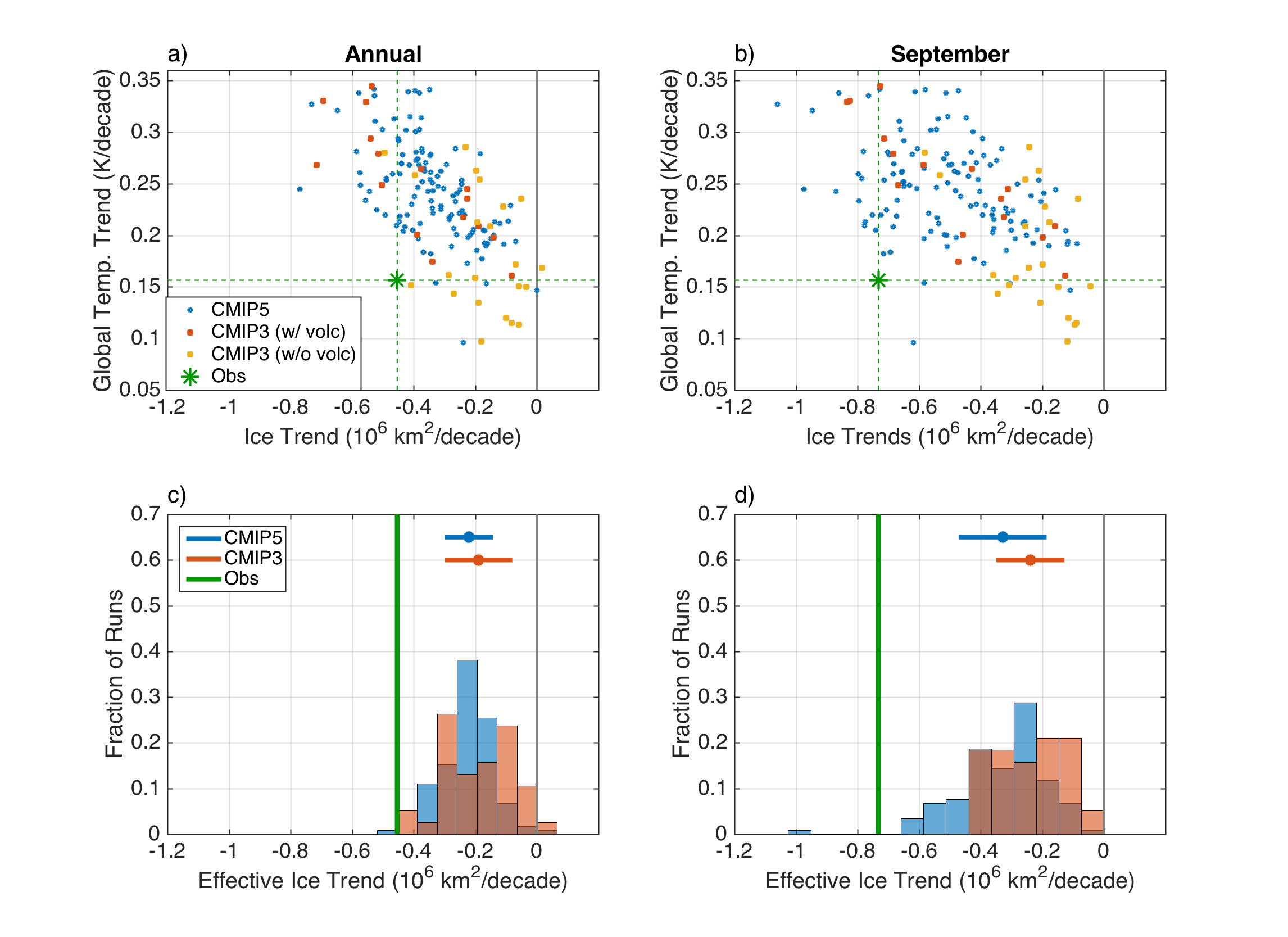}
 \caption{As in Figure 5, but using sea ice area rather than extent.}
\label{effTrendArea}
\end{figure*}